\documentstyle[aps,preprint]{revtex}
\input epsf
\tightenlines
\begin{document}

\title{On the Nature of the `Quasiparticle' Peak in the Angular Resolved
Spectrum of the Superconducting Underdoped Bi$_2$Sr$_2$CaCu$_2$O$_8$.}

\author{A.V. Rozhkov}

\address{
Center~for~Materials~Theory,
Department~of~Physics~and~Astronomy, Rutgers~University,
136~Frelinghuysen~Road,
Piscataway, NJ~08854,~USA
}

\maketitle

\begin{abstract}
We study the reconfiguration of the angular resolved
photoemission spectrum near $M$ point which occurs in
Bi$_2$Sr$_2$CaCu$_2$O$_8$ upon cooling below the
superconducting transition temperature.
Restricting our attention to the case of
underdoped samples we offer a phenomenological mechanism-independent argument
which explains the emergence of a peak in the spectrum in terms of the
normal state pseudogap. All of the basic experimental observations, 
including the peak
dispersion, `dip-and-hump' shape of the superconducting state
spectrum and appearance of the peak at the temperatures somewhat higher than
$T_c$, are naturally explained.
\end{abstract}
\hfill
\draft

\section{Introduction} 
A sharp peak in the angular resolved 
photoemission spectrum 
of Bi$_2$Sr$_2$CaCu$_2$O$_8$ (BSCCO) is the major subject of this paper 
\cite{loeser,arg,stan}. This peak 
emerges from an incoherent background at temperatures somewhat higher than
$T_c$ and persists all the way down to $T=0$. It is
located around $(\pi,0)$ (in the units of inverse lattice spacing) in 
the Brillouin zone. Substantial efforts have been invested into uncovering
the nature of this peak. It was assumed that this might provide us with an
important clue about a mechanism of the superconducting state.
In this paper we will argue that the peak is a 
consequence of normal state phenomenology. Our method will enable us to
explain qualitatively the most salient features of the peak behavior.

Unlike overdoped cuprates, 
it is believed that for the underdoped materials the normal state
single-particle Green's function can be qualitatively characterized by the
presence of the pseudogap -- a depletion of the spectral density $A_{n{\bf
k}}(\omega)=-2
{\rm Im\ }G_{n{\bf k}}(\omega)$ near $\omega=0$ for ${\bf k}$ in the
vicinity of $M$ point of the Brillouin zone
close to the Fermi surface (see fig.1b). The origin of the pseudogap is
unclear at the moment. Yet, our method does not require such knowledge.

The causality puts some constraints on the Green's function:
its real part can be 
found according to the Kramers transformation \cite{mahan}.
Then, for any reasonable choice of "pseudogapped" 
$A_{n\bf k}(\omega)$ the function 
$\left|{\rm Re} G_{n\bf k}(\omega)\right|$ 
has two maxima located approximately at the edges of the pseudogap (fig.1a).
We will show that those maxima are responsible for the 
emergence of the `quasiparticles' in the superconducting state.

There are several microscopical theories which explain the emergence of the
peak below $T_c$ \cite{hirsch,eschrig,erika}. 
In the approach presented by J.E. Hirsch
charge carriers (holes) are coupled to a bosonic bath. If this coupling
weakens with the growth of 
the local hole concentration the superconducting pairing driven
by the kinetic energy occurs: the kinetic energy benefits from the effective
reduction of the boson-hole coupling. Another consequence of this
boson-hole decoupling is the increase of the quasiparticle peak weight.
In the paper by M. Eschrig {\it et. al.} a model of the electrons interacting
with a magnetic resonance has been discussed. It is claimed that this model
reproduces correctly the most salient features of the photoemission 
spectrum. This paper was a development of a previous work by M. Norman,
H. Ding and collaborators \cite{norm,norm2}. 
The latter is important to us since it has some 
similarities with the present approach. In Ref. \cite{norm,norm2} the authors
concentrated their attention on a step-like feature in the electron
self-energy which could explain the shape of the spectrum. In order to
account for such a feature they postulated the existence of 
a collective mode below $T_c$. 
This mode has to disappear above $T_c$ for the model 
predictions to be consistent with the data. It is believed that this mode
and the magnetic resonance mentioned above are the same entity.
Many-body investigation of the phenomenon was undertaken
by E. Carlson {\it et. al.} They assume that
in the cuprate materials phase separation takes place.
It leads to the formation of stripes
with every stripe being 1D metallic conductor.
Using bosonization the authors demonstrated
that below $T_c$ the single-particle spectral
density develops a delta-function peak at the superconducting gap energy.

In contrast to these studies, we will deduce this peak without resorting to
microscopical considerations. It will be shown that
the peak in the single-particle
spectral density is a consequence of the analytical structure of the normal 
state Green's function.

\section{Analytical structure of the single electron propagator}

In this part of the paper we will calculate the spectral density in the 
superconducting state.
Let us derive first an auxiliary relation between the Green's function of an
electron and a hole. The spin-up electron retarded propagator is:
$
G^{\rm e} (t) = -i\Theta(t)\langle c_\uparrow(t)c_\uparrow^\dagger (0) 
+ c_\uparrow^\dagger (0)c_\uparrow(t)\rangle.
$
The propagator for a spin-up hole:
\begin{eqnarray}
G^{\rm h} (t) = -i\Theta(t)\langle c_\downarrow^\dagger (t) 
c_\downarrow^{\vphantom{\dagger}}(0) 
+ c_\downarrow^{\vphantom{\dagger}}(0)c_\downarrow^\dagger (t)
\rangle = \left( i \Theta(t) \langle c_\downarrow^\dagger (0) 
c_\downarrow^{\vphantom{\dagger}}(t) + c_\downarrow^{\vphantom{\dagger}} (t) 
c_\downarrow^\dagger(0)
\rangle \right)^* =\\ 
\left( i \Theta(t) \langle U_x^{-1}c_\downarrow^\dagger (0)U_x^{\vphantom{-1}} 
U_x^{-1}c_\downarrow^{\vphantom{\dagger}}(t)U_x^{\vphantom{-1}} + 
U_x^{-1}c_\downarrow^{\vphantom{\dagger}} (t)U_x^{\vphantom{-1}} U_x^{-1} 
c_\downarrow^\dagger(0)U_x^{\vphantom{-1}}
\rangle \right)^*=
\left(-G^{\rm e}(t)\right)^*.\nonumber
\end{eqnarray}
Here $U_x$ is the unitary rotation operator. 
It rotates the spin of the electron by $\pi$ around $x$ axis:
$
U_x^{-1}c_\alpha^{\vphantom{-1}} U_x^{\vphantom{-1}} = i\sigma^x_{\alpha\beta}
c_\beta^{\vphantom{x}}
$
and
$
U_x^{-1}c_\alpha^\dagger U_x^{\vphantom{-1}} =
-i\sigma^x_{\alpha\beta}
c_\beta^\dagger.
$
The ground state is invariant under the action of $U_x$.
After performing Fourier transformation for the last equality one arrives at:
\begin{equation}
G^{\rm h} (\omega) = -\left(G^{\rm e}(-\omega) \right)^*\label{rel1}.
\end{equation}

Assume now that the system at hand is in the normal state but not far from
the superconducting transition. 
Its retarded Green's function in Nambu representation can be written as:
\begin{equation}
{\hat G}_{n{\bf k}}(\omega)=\left(\matrix{G_{n{\bf k}}(\omega)&0\cr
				0&-G^*_{n{\bf -k}}(-\omega)\cr }\right)=
\left(\matrix{G_{n{\bf k}}(\omega)&0\cr
                                0&-G^*_{n{\bf k}}(-\omega)\cr }\right)
\end{equation}
In this form it
describes independent propagation of an electron and a hole. Spatial
inversion symmetry guarantees that $G_{\bf -k}=G_{\bf k}$. 

This
function $\hat G_n$ satisfies Dyson's equation:
\begin{equation}
(\omega-{\cal H}_0){\hat G}_{n{\bf k}}(\omega)={\hat 1}+
{\hat \Sigma}_{n{\bf k}} 
(\omega) {\hat G}_{n{\bf k}}(\omega).\label{DysonN}
\end{equation}
Here ${\hat \Sigma}_n$ is the retarded self-energy in the normal state:
\begin{equation}
{\hat \Sigma}_{n{\bf k}}(\omega) = \left(\matrix{\Sigma_{n{\bf k}}(\omega)
&0\cr 0&-\Sigma^*_{n{\bf k}}(-\omega)} \right).
\end{equation}
Next, we change, let's say, doping and drive the system superconducting. New
Green's function ${\hat G}_s$ satisfies an equation similar to 
(\ref{DysonN}) with new
self-energy ${\hat \Sigma}_{s{\bf k}}={\hat \Sigma}_{n{\bf k}} + 
{\hat \sigma_{\bf k}}$ where ${\hat \sigma}$ given by:
\begin{equation}
{\hat \sigma}_{\bf k}(\omega)=\left(\matrix{\mu_{\bf k}(\omega)&\sigma_{\bf
k}^{\cal A}(\omega)\cr (\sigma_{\bf k}^{\cal A}(-\omega))^*& 
-\left(\mu_{\bf k}(-\omega)\right)^*\cr}\right).
\end{equation}
Here the relation $\Sigma_{12}(\omega)=(\Sigma_{21}(-\omega))^*$ between two
off-diagonal elements of the Nambu self-energy has been used. It follows from
an equation $G_{12}(\omega) = G_{21}^*(-\omega)$ for anomalous propagators 
which, in turn, can be obtained
along the same lines as (\ref{rel1}). The point of this derivation is to
establish the analytical structure of the self-energy in the
superconducting state. This will be used in our discussion of a sum rule.

Combining the above equations one gets for ${\hat G}_s$ the expression:
\begin{equation}
{\hat G}_{s{\bf k}}=\left( {\hat G}^{-1}_{n{\bf k}} - {\hat \sigma}_{\bf k}
\right)^{-1}.
\end{equation}

\section{Green's function in the superconducting state}
The equations derived in the previous section are completely general. In
order to proceed further we need to 
make a simplifying assumption: near the Fermi surface 
electronic
$(\omega<0)$ and hole $(\omega>0)$ parts of the spectral function are
symmetric for small $|\omega|$. One can express that as:
\begin{equation}
A_{\bf k}(\omega)=A_{\bf k}(-\omega)\label{symm}
\end{equation}
for $|\omega|<\omega_g$ where $\omega_g$ is the size of the pseudogap.
Exact particle-hole symmetry is absent in the cuprates. 
We would like to argue,
however, that (\ref{symm}) holds at least approximately for ${\bf k}$
close to the Fermi surface. Firstly, the
tunneling data \cite{tunneling} shows that the pseudogap is symmetric 
with respect to voltage polarity change.
One might think that this tells little about validity of
(\ref{symm}) since the tunneling differential conductance measures ${\bf
k}$-integrated spectral density. However, Ding {\it et al.} (inset of
fig.1b of Ref. \cite{arg}) 
demonstrated that scanning tunneling microscope (STM) 
spectrum multiplied by the
Fermi distribution function at appropriate temperature coincides with the
angular-resolved spectral density at ${\bf k} = (\pi,0)$. This means that STM
spectrum equal to $A_{\bf k}(\omega)$ for $\omega<0$. It is natural
to expect that the spectral density coincide with the STM conductivity
for $\omega>0$ as well. Thus, the symmetry of the STM data suggests the
symmetry of the photoemission spectrum. Although, it is not our goal to
construct a theory of tunneling into cuprates we may speculate
that this agreement between the tunneling conductance and the photoemission
spectrum is not a coincidence but a consequence of the
inter-plane hopping matrix element ${\bf k}$-dependence: as discussed before
\cite{chasudand,andersen,xiapancoo} 
the inter-plane hopping matrix element is at
its maximum for the electrons with a momentum parallel to $(\pi,0)$ and it
is vanishes along Brillouin zone diagonal. Therefore, it is conceivable
that the most of the tunneling current is carried by the electrons near
$(\pi,0)$. Secondly, we would like to direct our attention toward the
discussion of Ref. \cite{randeria}. In the latter paper the angular
resolved spectrum at ${\bf k} = {\bf k}_{\rm F}$ was analyzed with the help
of a momentum distribution sum rule. Theoretically, 
it was proved that an expression
$1-\int_0^\infty d\omega \tanh (\omega/2T)(A_{{\bf k}_{\rm F}} (\omega) - 
A_{{\bf k}_{\rm F}} (-\omega))$ has to be temperature independent provided
that (\ref{symm}) is true for small $|\omega|$ \cite{caut1}. This expression
can be easily extracted from the data since it is proportional to the 
integral over $\omega$ of the 
photoemission spectrum. The spectrum for ${\bf k}$ near $(\pi,0)$ was 
integrated, and the integral showed no dependence on the temperature,
supporting the assumption of the symmetry. Finally, we would like to
mention two other papers \cite{norm3,norm4} where (\ref{symm}) was assumed
to be true. Thus, we believe we have enough evidence to treat 
(\ref{symm}) as a reasonable approximation for small $|\omega|$. The
structure of the background for $|\omega|>\omega_g$ is of little interest
to us since the positions of $|{\rm Re\ }G_n|$ maxima are insensitive to
the background. For the purpose of simplicity we will assume that the
background is also symmetric. That is, (\ref{symm}) is true for any
$\omega$. In such a case Green's function satisfies $G_{\bf k}(-\omega) =
-G_{\bf k}^*(\omega)$.

A consequence of the symmetry is a constraint on the 
self-energy in the superconducting phase:
\begin{equation}
{\hat \sigma}_{\bf k}(\omega)=\left(\matrix{\mu_{\bf k}(\omega)&\sigma_{\bf
k}^{\cal A}(\omega){\rm e}^{i\phi}\cr 
\sigma_{\bf k}^{\cal A}(\omega){\rm e}^{-i\phi}&
\mu_{\bf k}(\omega)\cr}\right),\label{sigma}
\end{equation}
where $\phi$ is a real number and
the functions $\mu$ and $\sigma^{\cal A}$ satisfy the following:
\begin{equation}
\mu_{\bf k}(-\omega)= - \mu_{\bf k}^*(\omega)\ \ {\rm and}\ \ 
\sigma_{\bf k}^{\cal A} (-\omega) = (\sigma_{\bf k}^{\cal A}(\omega))^*.
\end{equation}
In terms of these functions the superconducting spectral density equals to:
\begin{equation}
A_{s\bf k} = -2
{\rm Im\ }\left\{{ G_{n\bf k} - \left( G_{n\bf k} \right)^2 
\mu_{\bf k}
\over \left( 1 - G_{n\bf k}\mu_{\bf k}
\right)^2 - \left( \sigma_{\bf k}^{\cal A} \right)^2 \left(
G_{n\bf k}\right)^2}\right\}.\label{ASg}
\end{equation}
Let us simplify this equation. If the system is close to the transition
point then ${\hat \sigma}$ is small and the denominator of (\ref{ASg})
is close to unity everywhere except the edges
of the pseudogap. At those points as one can see from fig.1a
$({\rm Re\ }G_n)^2$ is big and one should be more
careful. If we are away from the Brillouin zone diagonal where 
$\sigma^{\cal A}$ is zero then
to the second order in $\sigma^{\cal A}$ and to the first order in $\mu=
{\cal O}(|\sigma^{\cal A}|^2)$ we find that the denominator is
equal to $1-(\sigma^{\cal A})^2(G_n)^2$. In this expression the terms 
like $G_n\mu$
have been dropped. Although, they are of the order of $|\sigma^{\cal A}|^2$ 
we presume that they are smaller than $({\rm Re\ }G_n)^2|\sigma^{\cal A}|^2$ 
since the latter is proportional 
to the second power of ${\rm Re\ }G_n$ which is big at the pseudogap edges.
This gives for the spectral function:
\begin{equation}
A_{s\bf k}(\omega) \simeq {A_{n\bf k}(\omega)\over 1-{\rm Re}\left\{\left(
\sigma^{\cal A}_{\bf k} (\omega)
\right)^2\right\} \left({\rm Re\ } G_{n \bf k}(\omega)\right)^2},\label{AS}
\end{equation}
where we omit $(G_n)^2\mu$ term in the numerator assuming that it is small
compare to $G_n$. The last formula is the key to understanding the
appearance of the `quasiparticles' below $T_c$. At the edges of the pseudogap
$({\rm Re\ }G_n)^2$ peaks up, the denominator gets smaller and the 
spectral function develops a sharp maximum. This is true provided that
${\rm Re}\left\{\left(\sigma^{\cal A}\right)^2\right\}$ is positive at the
edges of the pseudogap \cite{expl}.

\section{Spectrum properties} 
Now we would like to analyze (\ref{ASg}) and (\ref{AS}) to
obtain simple properties of the `quasiparticles' which can be compared against
the experiment.

First, it is clear that the position of the peak in the frequency domain is
determined by the pseudogap edge. Therefore, we expect to see strong
correlation between the size of the pseudogap and the frequency (binding
energy) position of the
peak. This is precisely the effect reported in \cite{loeser} where 
temperature dependence of the peak frequency was compared against that of
the pseudogap size.  Fig.2C of that paper shows that the peak position
traces the size of the pseudogap. Similar phenomena is seen for the peak
dispersion (binding energy vs. ${\bf k}$). As reported in \cite{mardesloe}
the pseudogap at $(\pi,0)$ is virtually ${\bf k}$-independent. This is
consistent with the weak dispersion of the peak itself \cite{disp}. 

In order to obtain the shape of the spectral function at ${\bf k}=(\pi,0)$
we performed simple numerical study.
We modeled $\mu,\ \sigma^{\cal A}$ and $A_n$ as
\begin{equation}
\begin{array}{c}
\mu=0\cr\cr
{\rm Re\ }\sigma^{\cal A}_{\bf k}
=\sigma^{\cal A}_0\exp(-(\omega/\omega^{\cal A})^2)
\cr\cr
A_{n\bf k} = A_0 \left(\arctan
\left(\left(\omega-\omega_g\right)/\omega_e\right) - \arctan 
\left(\left(\omega+\omega_g\right)/\omega_e\right)+\pi\right)
\exp\left(-\left|\omega/W\right|\right).\cr
\end{array}\label{model}
\end{equation}
The energy $\omega_e$ determines how
abrupt the edges of the pseudogap, $W$ mimics the bandwidth.
The constant $A_0$ has to be determined from the sum rule \cite{mahan}:
\begin{equation}
\int_{-\infty}^{+\infty} \frac{d\omega}{2\pi} A_{\bf k} (\omega) = 1.
\label{sumrule}
\end{equation}
The
parameter $\omega^{\cal A}$ describes the frequency dependence of the
anomalous self-energy. It is commonly put equal to infinity. This makes the
anomalous self-energy frequency independent. 
Below we will discuss both
cases of finite and infinite $\omega^{\cal A}$.

The imaginary part of $\sigma^{\cal A}$ and the real part of
the Green's function can be found with the help of numerical Kramers 
transformation. Then
the spectral function in the superconducting state is determined using 
(\ref{ASg}). The resultant spectra for different parameter values
are presented on fig.2 and fig.3. These two figures correspond to different
values of $\omega^{\cal A}$ ($\omega^{\cal A}/\omega_g=15$ for fig.2 and
$\omega^{\cal A}/\omega_g=7$ for fig.3). On each figure the top (bottom) 
row corresponds to $\sigma^{\cal A}/\omega_g = 25$
$(\sigma^{\cal A}/\omega_g = 15)$. The value of $\omega_e/\omega_g$ in the
left (right) column is 0.3 (0.1).
We see clear `dip-and-hump' structure of the spectrum for big values of
$\sigma^{\cal A}$. When this quantity is small we find only a tiny bump or
even a kink instead of a well-developed peak.
Such situation seems to be realized in Pb-doped
BSCO compound \cite{satkamnai}. This angular resolved
photoemission study revealed no peak in the spectrum (fig.3,
right panel of the latter reference). 
Authors claim that their energy resolution is high enough to
observe the peak should it be present in the spectrum.

By comparing different spectra from fig.2 and 3 we can discuss
the dependence of the spectrum shape on the model parameters.
Two parameters which affect the spectrum the most are 
$\sigma^{\cal A}/\omega_g$ and $\omega_e/\omega_g$.
The effect of $\sigma^{\cal A}$ is obvious: the
bigger the anomalous self-energy the higher the peak. The
ratio $\omega_e/\omega_g$ regulates the height of the peaks of ${\rm Re\ }
G_n$. This height is proportional to $-\ln(\omega_e/\omega_g)$.
Thus, the smaller this ratio the higher the peak in the
superconducting spectrum. The value of $\omega^{\cal A}$ specifies roughly
the position of the hump. For $\omega^{\cal A} = \infty$ the hump is absent
from the spectrum -- it is moved to infinite frequency. 

Another interesting property of our superconducting state spectrum is its
compliance with the sum rule (\ref{sumrule}). Mathematical proof of this
fact is given in Appendix. Visually, one can notice from figures 2 and 3
that the depletion of the spectral weight in the dip is roughly
compensated by its enhancement at the hump. Experimentally, however, it is
seen that the weight from the dip goes into the peak. The reason for this
discrepancy is the lack of exact knowledge about $\hat \sigma$. Particular,
we are
unable to identify any first principal based constraint which would provide
us with any information about $\mu$. To keep our consideration as simple as
possible we chose to put $\mu$ equal to zero.

As we stressed the equality (\ref{symm}) is only an approximation. Thus,
an interesting question worth discussing is how deviations
from (\ref{symm}) affect the peak in the superconducting state. To address
this issue we
model the electron spectral function as
\begin{equation}
A_{n\bf k}^{\rm e} (\omega) = A_{n\bf k}(\omega+\Delta\omega),
\end{equation}
where $A_{n\bf k}(\omega)$ is given by (\ref{model}).
In this case the pseudogap for electrons is bigger then the pseudogap
for holes by the amount of $2\Delta\omega$. The spectral function can be
found according to 
\begin{equation}
A_{s\bf k}^{\rm e} = -2
{\rm Im\ }\left\{{ G_{n\bf k}^{\rm e}
\over 1- \left( \sigma_{\bf k}^{\cal A} \right)^2 
G_{n\bf k}^{\rm h}G_{n\bf k}^{\rm e}}\right\}.\label{ASeh}
\end{equation}
The resulting function for $\Delta\omega=0.2\omega_g$ and
$\sigma^{\cal A}(\omega) \equiv 25\omega_g$ is plotted on fig.4. 
As one can see the only
effect of the asymmetry is uneven hight of the peaks for positive and
negative frequencies. The reason for this becomes clear if one notice that
(\ref{ASeh}) may be approximated in the manner similar to (\ref{AS}):
\begin{equation}
A_{s\bf k}^{\rm e} \simeq {A_{n\bf k}^{\rm e}
\over 1 - \left( \sigma_{\bf k}^{\cal A} \right)^2 
{\rm Re\ }\left\{
G_{n\bf k}^{\rm h}G_{n\bf k}^{\rm e}\right\}}.\label{ASehapp}
\end{equation}
In order to demonstrate the quality of this approximation we plotted 
this expression on fig.4.
The denominator in (\ref{ASehapp}) is symmetric with respect to the
frequency sign change. It has two minima located approximately at
$\pm\omega_g$. Since $A^{\rm e}_n(\omega_g)>A^{\rm e}_n(-\omega_g)$ the
peak at $\omega<0$ is smaller. It is now clear that the superconducting
state spectrum is rather robust with respect to small deviations from
(\ref{symm}).

To understand why the peak appears at the temperatures higher than $T_c$ it
is enough to imagine that in the cuprates there is a temperature window
$T_{\rm local}>T>T_c$ where the phase coherence exists over some finite 
time scale 
$\tau(T),\ \tau\rightarrow\infty$ as $T\rightarrow T_c$. Experimental
evidence of this was produced by \cite{corson}. 
Speaking more technically, in this temperature window
one can use (\ref{sigma}) with non-zero $\sigma^{\cal A}$ and
fixed $\phi$ to
describe the single-particle dynamics on the time scale less than $\tau$.
For time periods bigger than $\tau$ the dynamics of the phase $\phi$ cannot
be neglected anymore. If $\tau$ is big enough one can simply average the
propagator over the phase fluctuations.
For infinitely long time the anomalous propagator 
always vanishes above $T_c$ due to the phase 
dynamics. However, the peak in the photoemission spectrum does not
vanish since, as it follows from (\ref{ASg}), it is independent of $\phi$.

\section{Discussion} 
We study the `quasiparticle' peak in the
superconducting state of the underdoped BSCCO.
The main conclusion of the present work is that the peak 
can be explained on the basis of the normal state
phenomenology. Our analysis is based on 
three premises. First, we assume that the single particle spectrum has a 
pseudogap (fig.1b). 
The most important consequence of that is the peaks of the
function $|{\rm Re\ }G_n (\omega)|$ at the edges of the pseudogap. Next, we
assume that there is approximate symmetry between positive-frequency and 
negative-frequency parts of the spectrum (\ref{symm}). This guarantees that 
the function ${\rm -Re\ }\{G_n(\omega)G_n(-\omega)\}$ has only one maximum at 
$\omega<0$. We also shown that some violation of (\ref{symm}) is not fatal
for the qualitative structure of the superconducting state spectrum.
Finally, there is the third assumption we made implicitly: the
single-particle anomalous self-energy $\sigma^{\cal A}(\omega)$ does not go
to zero at the edges of the pseudogap. If $\sigma^{\cal A}$ does go to zero
then the peak in the superconducting phase could be
completely suppressed (see (\ref{AS})).

In order to explain the emergence of the peak at $T>T_c$ we accepted another
postulate. It was agreed that (\ref{sigma}) local order parameter is
established and it is permissible to use (\ref{sigma}) with fluctuating
$\phi$ to describe the
single-particle dynamics above $T_c$.

The fact that the peak can be derived from the normal state properties
suggests that it is not a fundamental object. Moreover, depending on the
circumstances it may be absent or replaced by a less prominent feature
\cite{satkamnai}.

It is also can be surmised that this phenomena may not be limited to the
cuprate superconductors only. The peak could be present in the
superconducting state of any material which has a pseudogap in the normal
state.

We did not try to fit any experimental data by our curves from fig.2 and
fig.3. It is not possible owing to the lack of knowledge about exact form
of functions (\ref{model}). We also have no independent information about
the values of $\omega_e,\ \omega^{\cal A},\ \sigma^{\cal A}_0$.
Another argument against feasibility of the spectrum fit is a problem
with the experimental data themselves. It is known that the photoemission
spectrum is a convolution of the single-electron Green's function and a
photoemission matrix element. It has been assumed that this matrix element
is a constant, at least in a relevant range of parameters. Thus, relative
intensities of different features of
the spectrum were believed to be free of any influence extrinsic to 
the electron dynamics.
However, it was shown recently that the matrix element
has very non-trivial dependence on the photon energy and the binding energy
of the photoelectrons \cite{borisenko}. In such a situation the significance
of an intensity fit is substantially devaluated.

To conclude, we study the peak in the photoemission spectrum of BSCCO. It
was shown that this peak is a consequence of the normal state pseudogap.
We derived the most basic properties of the peak and demonstrated that they
are in qualitative agreement with the experimental data available.

\section{Acknoledgements} 
Author is most grateful to D. Basov, A.J.
Millis and J.E. Hirsch for help and discussions.

\section{Appendix}

In this Appendix we will prove that under rather mild assumptions on
$G_n$ and $\sigma^{\cal A}$ the spectral function (\ref{ASg}) satisfies
(\ref{sumrule}). In our derivation we will assume that $\mu\equiv 0$. This
limitation is not crucial.

An essential part of our proof is the analyticity of both $G_n(\omega)$ and
$\sigma^{\cal A}(\omega)$ for complex values of $\omega, {\rm Im\ }\omega >
0$ (upper half-plane). 
First, we establish asymptotical behavior of $G_n$ at large 
$|\omega|$. If $|\omega| \gg W$ the details of the spectral function 
$A(\omega)$ are
irrelevant and it can be viewed as a delta function times $2\pi$. This means
that at large $|\omega|,\ {\rm Im\ }\omega>0$ 
\begin{equation}
G_n(\omega) = \int_{-\infty}^{+\infty} {d\omega'\over 2\pi} {A_n(\omega')
\over \omega - \omega'} \simeq {1\over\omega}.
\end{equation}
We see, that due to the sum rule (\ref{sumrule}) the residue of $G_n$ at
infinity is exactly unity.
Next, for any finite $\omega^{\cal A}$ the same reasoning can be applied
to $\sigma^{\cal A}$ with the result:
\begin{equation}
\sigma^{\cal A}(\omega) = {\cal O}(|\omega|^{-1})
\end{equation}
for large $|\omega|$. We do not try to determine the residue of
$\sigma^{\cal A}$ at intinity since it is of no interst to us. 
The asymptotical behavior combined with the absence of any singularity in
the upper half-plane of complex $\omega$ implies that both $G_n$ and
$\sigma^{\cal A}$ are bounded:
\begin{eqnarray}
|\sigma^{\cal A} (\omega)| < C_\sigma,\\
|G_n(\omega)| < C_G.
\end{eqnarray}
where $C_{\sigma,G}$ are some real positive constants. 
Now we are in position to prove that:
\begin{eqnarray}
{\rm Im\ } \int_{-\infty}^{+\infty} d\omega \left(G_s(\omega)-G_n(\omega)
\right)=0,
\label{sumrule2}\\
G_s(\omega)={G_n(\omega) \over
1-(\sigma^{\cal A}(\omega))^2 (G_n(\omega))^2},
\end{eqnarray}
provided that the product $C_G C_\sigma < 1$. This inequality
guarantees that the denominator of $G_s(\omega)$ never goes to zero.
Therefore, the integrand of (\ref{sumrule2}) has no singularities for ${\rm
Im\ }\omega>0$. This means that we can transform the integration contour
into a semi-circle in the upper half-plane.
The integral over the semi-circle of infinite
radius equals to the residue at infinity times $(-\pi)$. We already know
that this residue is unity for $G_n$. It is also unity for $G_s$: at large
$|\omega|$ the denominator is non-singular and approaches 1, thus, the
leading asymptotic behavior of $G_s$ is the same as that of $G_n$. 
The residue of the integrand is zero which proves (\ref{sumrule2}).

\begin{figure} [!t]
\centering
\leavevmode
\epsfxsize=8cm
\epsfysize=8cm
\epsfbox[18 144 592 718] {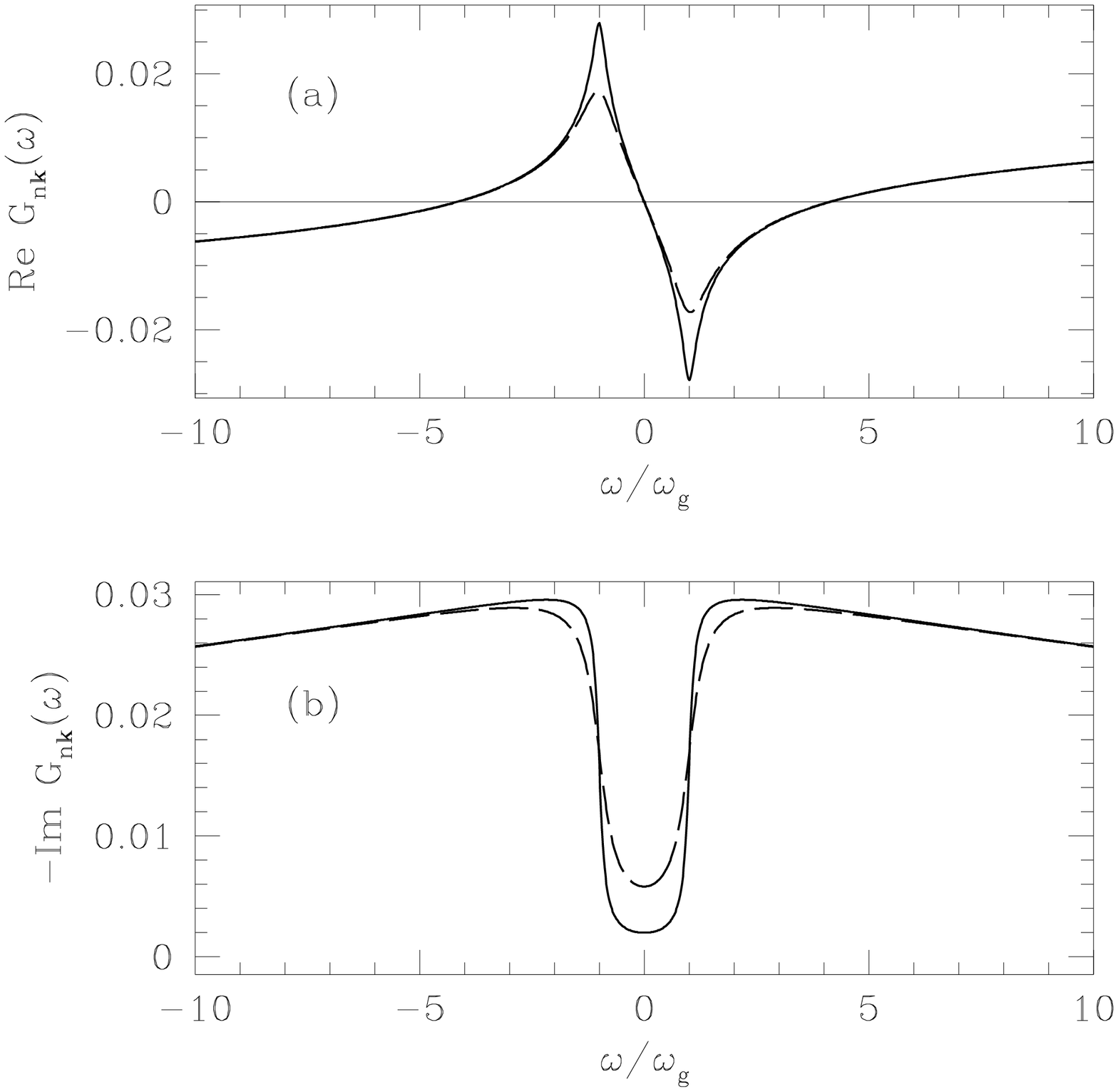}
\caption[]
{\label{fig1} Qualitative structure of the normal state Green's function
for different values of $\omega_e/\omega_g$. 
The real part of $G_n(\omega)$ (panel (a)) can be obtained by applying 
Kramers transformation 
to the imaginary part, panel (b). The solid line (dash-line) corresponds to
$\omega_e/\omega_g=0.1$ ($\omega_e/\omega_g=0.3$).}
\end{figure}

\begin{figure} [!t]
\centering
\leavevmode
\epsfxsize=8cm
\epsfysize=8cm
\epsfbox[18 144 592 718] {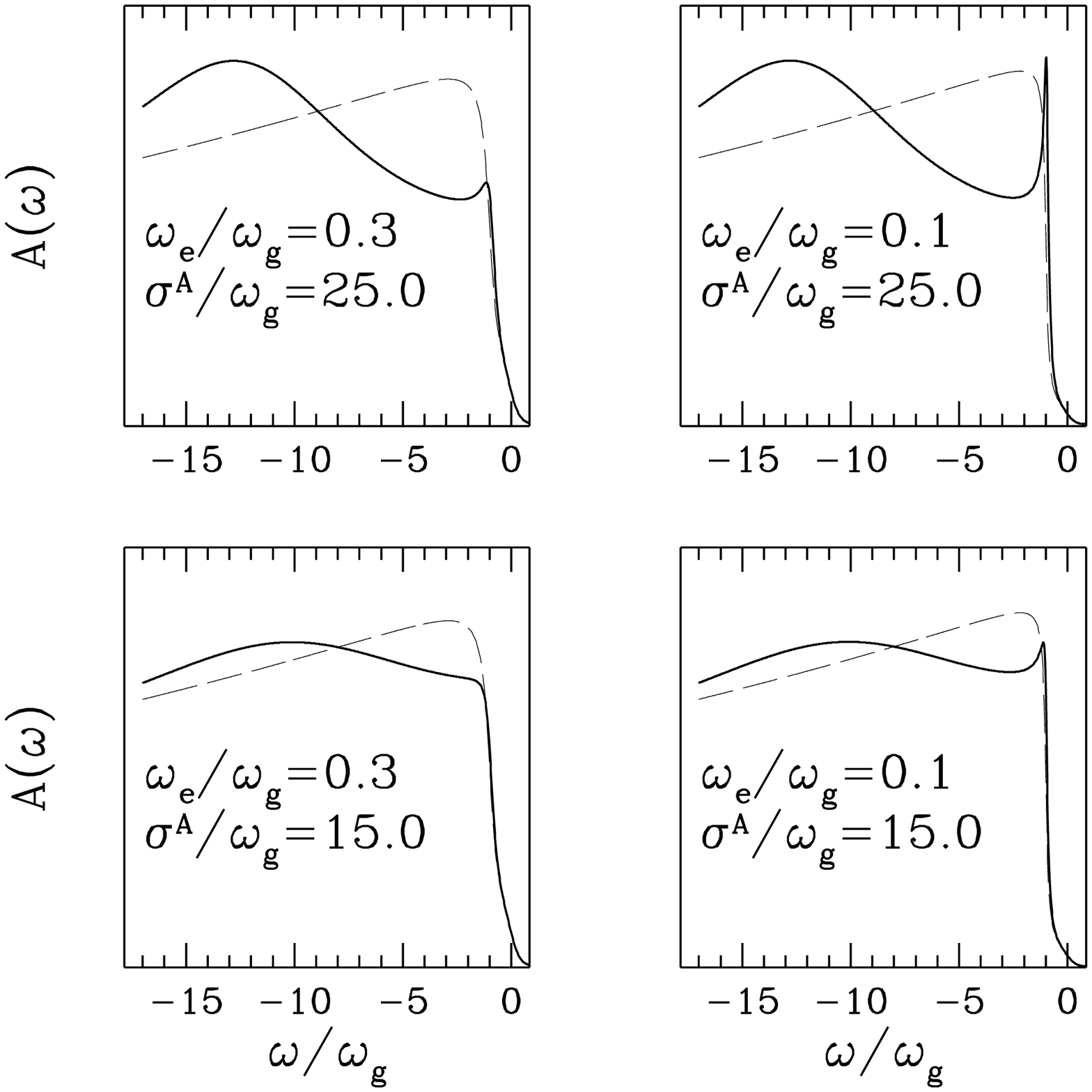}
\caption[]
{\label{fig2} Superconducting state spectral function ($A_s(\omega)\ vs.\
\omega/\omega_g$) for $W/\omega_g=50,\
\omega^{\cal A}/\omega_g=15$ and different values of 
$\sigma^{\cal A}/\omega_g$ and
$\omega_e/\omega_g$. The left (right) column corresponds to
$\omega_e/\omega_g = 0.3$ $(\omega_e/\omega_g = 0.1)$. The top (bottom) row
corresponds to $\sigma^{\cal A}/\omega_g = 25$ 
$(\sigma^{\cal A}/\omega_g = 15)$. The normal state spectrum is given in a
dash-line. The spectral functions are multiplied by the Fermi distribution
function with $T=0.2\omega_g$.
}
\end{figure}

\begin{figure} [!t]
\centering
\leavevmode
\epsfxsize=8cm
\epsfysize=8cm
\epsfbox[18 144 592 718] {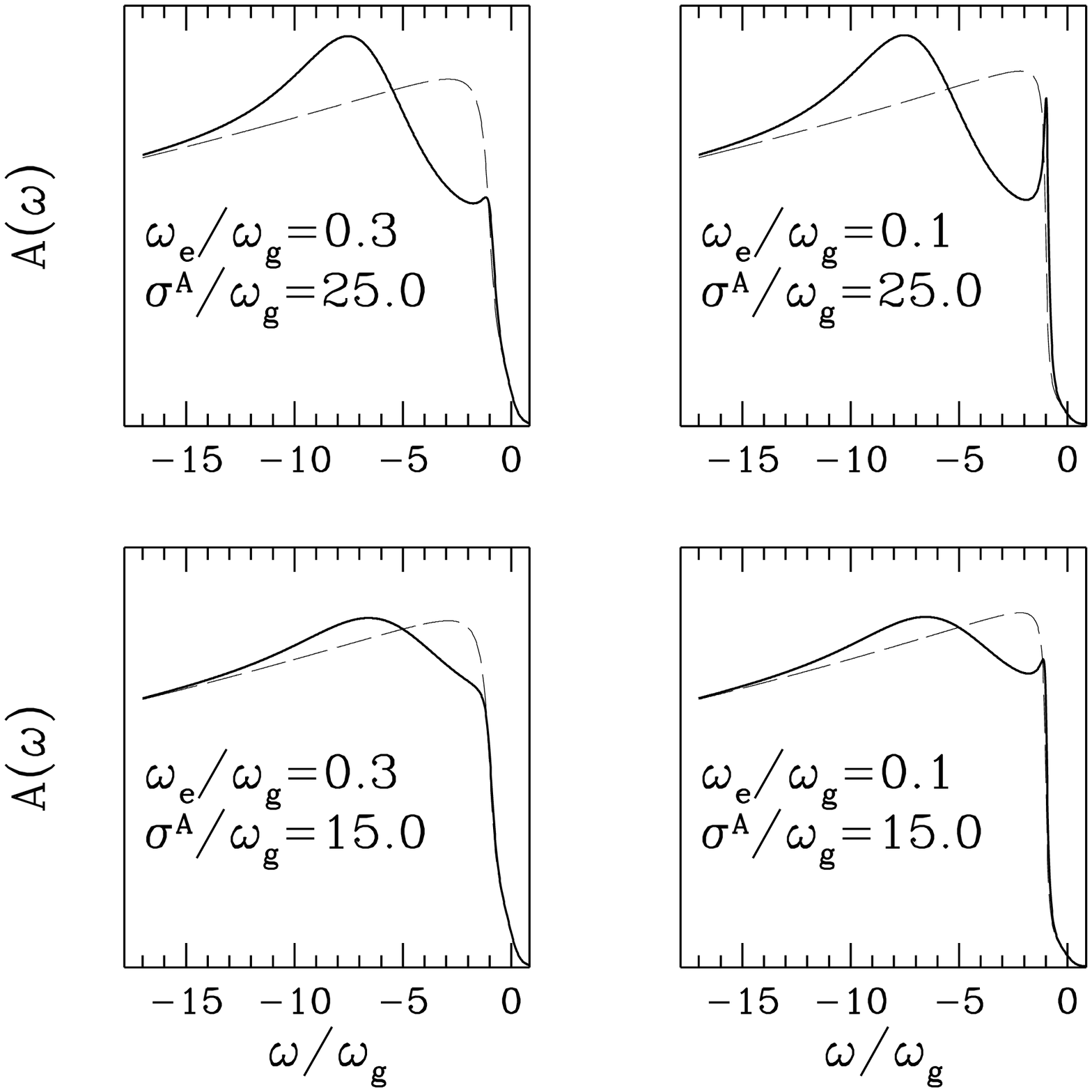}
\caption[]
{\label{fig3} Superconducting state spectral function ($A_s(\omega)\ vs.\
\omega/\omega_g$) for $W/\omega_g=50,\
\omega^{\cal A}/\omega_g=7$ and different values of 
$\sigma^{\cal A}/\omega_g$ and
$\omega_e/\omega_g$. The left (right) column corresponds to
$\omega_e/\omega_g = 0.3$ $(\omega_e/\omega_g = 0.1)$. The top (bottom) row
corresponds to $\sigma^{\cal A}/\omega_g = 25$ 
$(\sigma^{\cal A}/\omega_g = 15)$. The normal state spectrum is given in a
dash-line. The spectral functions are multiplied by the Fermi distribution
function with $T=0.2\omega_g$.
}
\end{figure}

\begin{figure} [!t]
\centering
\leavevmode
\epsfxsize=8cm
\epsfysize=8cm
\epsfbox[18 144 592 718] {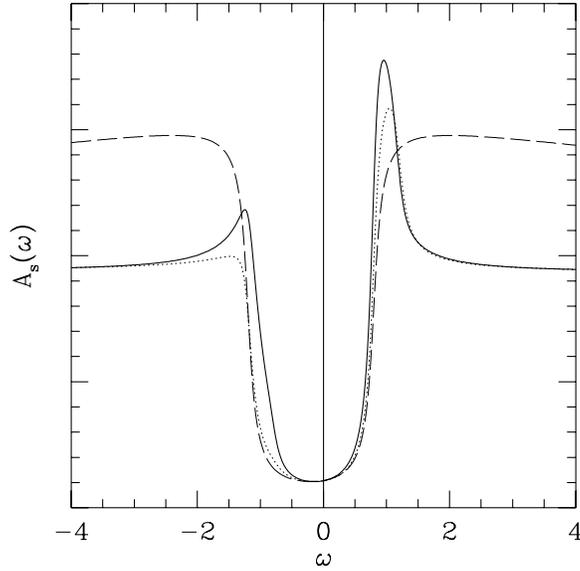}
\caption[]
{\label{fig4} The effect of weak particle-hole assymetry on the
superconducting spectral function. Normal state spectrum is shifted to the
left by amount of $\Delta\omega = 0.2\omega_g$. The values of parameters are:
$W/\omega_g=50$, $\omega^{\cal A} = \infty$, 
$\omega_e/\omega_g = 0.1$,
$\sigma^{\cal A}/\omega_g = 25$
The normal state spectrum is given in a
dash-line, the approximation (\ref{ASehapp}) is plotted in a dotted line.
}
\end{figure}


\begin{thebibliography}{99}
\bibitem{subscript} We use subscript `n' (`s') to denote normal
(superconducting) state quantities.
\bibitem{caut1} Unfortunately, the authors of \cite{randeria} did not
specify what is the frequency above which the symmetry of the spectrum is
violated.
\bibitem{expl} If $\sigma^{\cal A}$ is a frequency-independent quantity
then its imaginary part zero and, therefore, ${\rm Re\ }((\sigma^{\cal 
A})^2) = (\sigma^{\cal A})^2 > 0$.
Even if $\sigma^{\cal A}$ has some $\omega$-dependence 
${\rm Re\ }(\sigma^{\cal A})^2$ is still positive in the range of interest
provided that this dependence is a weak one.
\bibitem{disp} The author was unable to find any work done on the peak
dispersion for the underdoped BSCCO. However, the data for the optimally
doped and overdoped samples are available. For example, Ref. \cite{norm}
shows that the dispersion curve of the peak for optimally doped sample is
flat. Since most of the peak properties does not drastically change with
doping we expect to have no dispersion on the underdoped side as well.
\end{thebibliography}
\end{document}